\documentclass[a4paper,11pt]{article}
\usepackage{pos}
\usepackage[utf8]{inputenc}
\usepackage{amsmath}
\usepackage{float}
\usepackage{tabulary}
\usepackage{empheq}
\usepackage{tabu}

\newcommand{\code}[1]{\texttt{#1}}
\DeclareUnicodeCharacter{2212}{-}

\title{A study of dark matter-dark energy interaction under the
DESI DR2 data constraint}

\author[a]{Amin Aboubrahim}
\author*[b]{Pran Nath}

\affiliation[a]{Department of Physics, University of Hartford, \\
200 Bloomfield Avenue, West Hartford, CT 06117, U.S.A.}

\affiliation[b]{Department of Physics, Northeastern University, \\
111 Forsyth Street, Boston, MA 02115-5000, U.S.A.}

\emailAdd{abouibrah@hartford.edu}
\emailAdd{p.nath@northeastern.edu}

\abstract{While $\Lambda$CDM provides a good fit to cosmological data, it fails to address many of the outstanding questions in contemporary cosmology. Chief among these are the Hubble tension and the apparent dynamical nature of dark energy as inferred from the recent DESI DR2 analysis. In this work, we analyze a field-theoretic description of cosmology where both dark energy and dark matter are interacting spin zero fields. We give a thorough study of a wide range of the interaction strength and demonstrate the effect on the dark energy equation of state and the Hubble tension.  Using the recent cosmological data, we extract constraints on cosmological parameters including the free parameters of the model. }

\FullConference{Proceedings of the Corfu Summer Institute 2025 "Tensions in Cosmology 2025 Conference" (CORFU2025)\
September 2--8, 2025\\
Corfu, Greece\\}

\begin{document}
\maketitle

\section{Introduction} \label{sec:int}

The standard $\Lambda$CDM model successfully explains much cosmological data, but several tensions have emerged, most notably in the evaluation of the Hubble constant $H_0$ (for a recent review, see ref.~\cite{CosmoVerseNetwork:2025alb} and references therein). A common approach which may provide a solution is to model dark matter-dark energy (DM-DE) interactions through source terms in the continuity equations, but these prescriptions are not derived from a consistent field-theoretic framework. Additional discrepancies have also appeared in measurements of the dark energy equation of state (EoS). Recent DESI results~\cite{DESI:2024mwx,DESI:2025zgx} favor a time-varying EoS, which deviates from the $\Lambda$CDM prediction of $w = -1$. Quintessence models (particularly thawing and freezing types) provide alternative frameworks for generating such dynamical behavior without invoking the CPL parameterization~\cite{Chevallier:2000qy,Linder:2002et} which leads to phantom crossing.

This note is based on ref.~\cite{Aboubrahim:2024cyk} which examined whether a field-theoretic model of cosmology with DM-DE interaction can produce a dynamical EoS in agreement with DESI DR2, along with addressing the $H_0$ tension. The analysis avoids phenomenological continuity equations and instead uses a fully field-theoretic formulation that enforces energy conservation automatically. Within this framework, DM-DE coupling leads to rich EoS behavior: strong coupling can transform quintessence from thawing to scaling freezing, while weak coupling preserves the thawing regime. By fitting the EoS in both cases, the study evaluates how DM-DE interaction affects consistency with DESI
DR2 data and finds that the couplings influence the results, although the DESI DR2 data points to a weak DM-DE interaction.

\section{A model of cosmology with $n$ interacting fields}

We give a brief overview of the theory comprising of $n$ interacting spin zero fields, $\phi_i$, in the context of cosmology. This represents a generalization of the two-field case~\cite{Aboubrahim:2024spa,Nath:2025gzm}. The action of this theory, which we call QCDM, is given by
\begin{align}
 &S_{\rm QCDM}=\int \text{d}^4 x\sqrt{-g}
  \left[\frac{1}{16\pi G}R+ \sum_i\frac{1}{2}\phi_i^{,\mu}\phi_{i,\mu}-
  V\Big(\{\phi_i\},\{\phi_j\}\Big)\right],
\end{align}
where the potential is
\begin{align}
  &V=\sum_i^n V_i(\phi_i)+ V_{\rm int}(\{\phi_i\},\{\phi_j\})\,, \\
 &V_{\rm int}(\{\phi_i\},\{\phi_j\}) =\frac{1}{2} \sum_{i\neq j}\sum_jV_{ij}(\phi_i,\phi_j)\,.
\label{n2}    
\end{align}
The quantity $V_{ij}(\phi_i,\phi_j)$ is the interaction potential between the fields $\phi_i$ and $\phi_j$. The Klein-Gordon equation for the field $\phi_i$ is given by
\begin{equation}
\phi_i^{\prime\prime}+2\mathcal{H}\phi_i^\prime+a^2\left(V_{i,\phi_i}+
  \frac{1}{2} \sum_{j\neq i} (V_{ij}+ V_{ji})_{,\phi_i}\right)=0\,,
    \label{n3}
\end{equation}
where $V_{ij,\phi_i}\equiv \partial_{\phi_i}V_{ij}$, and  $\mathcal{H}=a^\prime/a$ is the conformal Hubble parameter. With the assumption $V_{ji}=V_{ij}$, the
corresponding continuity equations are then given by
\begin{align}
\label{n4a}
   & \rho^\prime_i+3\mathcal{H}(1+w_i)\rho_i= Q_i\,,\\
    &Q_i= \sum_{j\neq i}V_{ij,\phi_j}(\phi_i,\phi_j)\phi_j^\prime\,,
        \label{n4b}    
\end{align}
where $\sum_i Q_i\neq 0$ and the imposition of $\sum_i Q_i=0$ implies $Q_i=0$, i.e., no interaction at all. Further, the energy density $\rho_i$ and the pressure $p_i$ for the 
energy density of field $\phi_i$ are given by 
\begin{equation}
    \rho_i(p_i)=T_i\pm V_i(\phi_i)\pm\sum_{j\neq i}V_{ij}(\phi_i,\phi_j)\,,\
        \label{n6}    
\end{equation}
where $T_i$ is the kinetic energy of field $\phi_i$, the $+$($-$) signs are for the cases $\rho_i$ ($p_i$),
and $w_i= p_i/\rho_i$ is the equation of state.
The total energy density is then defined by
\begin{equation}
    \rho=\sum_i \rho_i-\sum_{i<j}V_{ij}(\phi_i,\phi_j)\,,
        \label{n7}
        \end{equation}
where the last term on the right hand side is included to ensure no double counting. The observed relic density for the particle $i$ is given by 
\begin{align}
  \label{n8a}
   \Omega_{0i}&=\frac{\rho_i}{\rho_{0,\rm crit}}(1-\epsilon_i), 
   ~~~~~~\sum_i \Omega_{0i}=1,~~~~~~
\epsilon_i=\frac{\sum_{j\neq i} V_{ij}}{
 \sum_j \rho_j}.
\end{align}
Eq.~(\ref{n4a}) represents  a consistent set of continuity equations for the case of $n$ number of interacting fields, where the energy density $\rho_i$ corresponds to the field $\phi_i$.

\section{Background and perturbation equations of QCDM}

In this section, we summarize the evolution of DM and DE fields in a flat FLRW universe. In the interacting quintessence-DM (QCDM) framework, the DM field $\chi$ and DE field $\phi$ evolve under the potentials
\begin{align}
V_1(\chi)&=\frac{1}{2}m_\chi^2\chi^2, \\
V_2(\phi)&=\mu^4\left[1+\cos\left(\frac{\phi}{F}\right)\right],
\end{align}
and interact through
\begin{align}
V_{\rm int}(\phi,\chi)=\frac{\lambda}{2}\chi^2\phi^2.
\end{align}
Perturbations are introduced as $\chi\to\chi+\chi_1$ and $\phi\to\phi+\phi_1$, with the synchronous gauge line element $\mathrm{d}s^2=a^2(\tau)[-\mathrm{d}\tau^2+(\delta_{ij}+h_{ij})\mathrm{d}x^i\mathrm{d}x^j]$.

The background fields satisfy the Klein-Gordon equations
\begin{align}
\label{kgc0}
\chi^{\prime\prime}+2\mathcal{H}\chi^\prime+a^2(V_{1,\chi}+V_{12,\chi})&=0\,, \\
\phi^{\prime\prime}+2\mathcal{H}\phi^\prime+a^2(V_{2,\phi}+V_{12,\phi})&=0\,.
\label{kgp0}
\end{align}
Because the DM field oscillates rapidly when $\mathcal{H}/m_\chi\ll 1$~\cite{Turner:1983he,Urena-Lopez:2015gur}, it is convenient to redefine the DM energy density as $\tilde{\rho}_\chi=\rho_\chi-V_{12}$ and introduce the modified fraction
\begin{align}
\tilde{\Omega}_\chi=\frac{\kappa^2}{6\mathcal{H}^2}\Big(\chi^{\prime 2}_0+2a^2V_1(\chi_0)\Big)\,,
\end{align}
along with new dimensionless variables~\cite{Copeland:1997et,Garcia-Arroyo:2024tqq}
\begin{align}
&\tilde{\Omega}_\chi^{1/2}\sin\left(\frac{\theta}{2}\right)=\frac{\kappa \chi^\prime}{\sqrt{6}\mathcal{H}}, \\
&\tilde{\Omega}_\chi^{1/2}\cos\left(\frac{\theta}{2}\right)=\frac{\kappa a V^{1/2}_{1}}{\sqrt{3}\mathcal{H}}, \\
&y=-\frac{2\sqrt{2}\,a}{\mathcal{H}}\partial_\chi V^{1/2}_{1}.
\end{align}
These variables absorb the rapid oscillations into $\theta$, allowing the DM KG equation to be rewritten as three first-order differential equations:
\begin{align}
\Omega^\prime_\chi&=3\mathcal{H}\Omega_\chi(w_T-w_\chi) 
+\frac{\kappa^2 a^2}{3\mathcal{H}^2}\Big[\mathcal{H}(1+3w_\chi)V_{12}-\chi^\prime V_{12,\chi}\Big]\,, \\
\theta^\prime&=-3\mathcal{H}\sin\theta+\mathcal{H}y 
-\frac{\kappa^2 a^2}{3\mathcal{H}^2\tilde{\Omega}_\chi}\Big(2\mathcal{H} V_{12}+\chi^\prime V_{12,\chi}\Big)\cot\frac{\theta}{2}\,, \\
y^\prime&=\frac{3}{2}{\cal H}(1+w_T)y\,.
\end{align}
In this parametrization, the DM equation of state becomes $w_\chi=-\cos\theta$,
which oscillates between $+1$ and $-1$, yielding a vanishing time average, consistent with CDM. The interaction term is omitted from $\tilde{p}_\chi$ and $\tilde{\rho}_\chi$ but reappears in the evolution equations.

Linear perturbations obey the KG equations
\begin{align}
\label{kgp1}
\phi_1^{\prime\prime}+2\mathcal{H}\phi_1^\prime+(k^2+a^2V_{,\phi\phi})\phi_1+a^2V_{,\phi\chi}\chi_1+\frac{1}{2}h^\prime\phi^\prime&=0\,, \\
\chi_1^{\prime\prime}+2\mathcal{H}\chi_1^\prime+(k^2+a^2V_{,\chi\chi})\chi_1+a^2V_{,\chi\phi}\phi_1+\frac{1}{2}h^\prime\chi^\prime&=0\,,
\label{kgc1}
\end{align}
leading to energy density and pressure perturbations
\begin{align}
\delta\rho_\phi&=\frac{1}{a^2}\phi^\prime\phi_1^\prime+({V}_2+{V}_{12})_{,\phi}\phi_1+{V}_{12,\chi}\chi_1\,, \\
\delta p_\phi&=\frac{1}{a^2}\phi^\prime\phi_1^\prime-({V}_2+{V}_{12})_{,\phi}\phi_1-{V}_{12,\chi}\chi_1\,, \\
\delta\rho_\chi&=\frac{1}{a^2}\chi^\prime\chi_1^\prime+({V}_1+{V}_{12})_{,\chi}\chi_1+{V}_{12,\phi}\phi_1\,, \\
\delta p_\chi&=\frac{1}{a^2}\chi^\prime\chi_1^\prime-({V}_1+{V}_{12})_{,\chi}\chi_1-{V}_{12,\phi}\phi_1\,.
\end{align}
The velocity divergences, $\Theta=ik^i v_i$, are
\begin{align}
(\rho_\phi+p_\phi)\Theta_\phi&=\frac{k^2}{a^2}\phi^\prime\phi_1\,, \\
(\rho_\chi+p_\chi)\Theta_\chi&=\frac{k^2}{a^2}\chi^\prime\chi_1\,,
\end{align}
and the density contrast is defined as $\delta=\delta\rho/\rho$.

\section{Phenomenological study of QCDM}

The model is implemented in the Boltzmann equation solver \code{CLASS}~\cite{Blas:2011rf} which evolves the background and perturbation equations from $a_{\rm ini}=10^{-14}$ to $a_0=1$ (today). The input parameters of the model are $\mu^4, F, \lambda, \phi_{\rm ini}, \phi^\prime_{\rm ini}, \chi_{\rm ini}, \chi^\prime_{\rm ini}$. The DE field velocity is chosen as $\phi^\prime_{\rm ini}=10^{-3}$ and an estimate of $\mu^4$ is determined by minimizing the DE potential so that $\mu^4=\frac{3}{2}H_0^2\Omega_{\phi 0}$, where $\Omega_{\phi 0}$ is today's DE density fraction. This value serves as an initial estimate and we modify $\mu^4$ accordingly in order to achieve a consistent cosmology. 

\begin{figure}[H]
\begin{centering}
\includegraphics[width=0.49\linewidth]{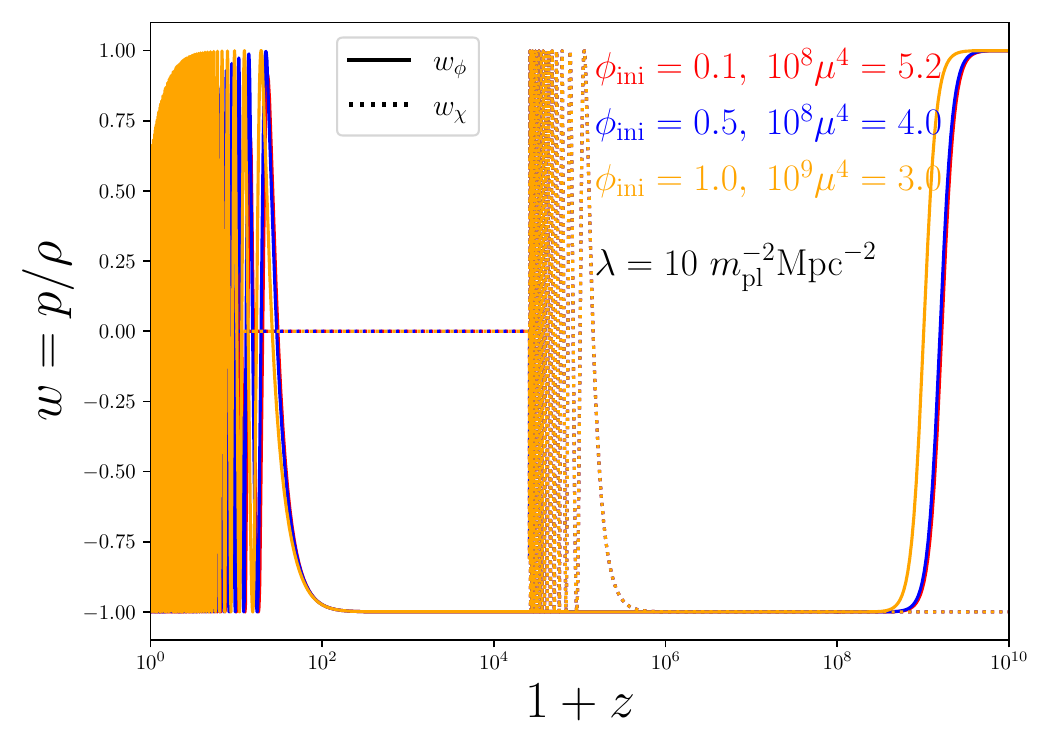}
\includegraphics[width=0.49\textwidth]{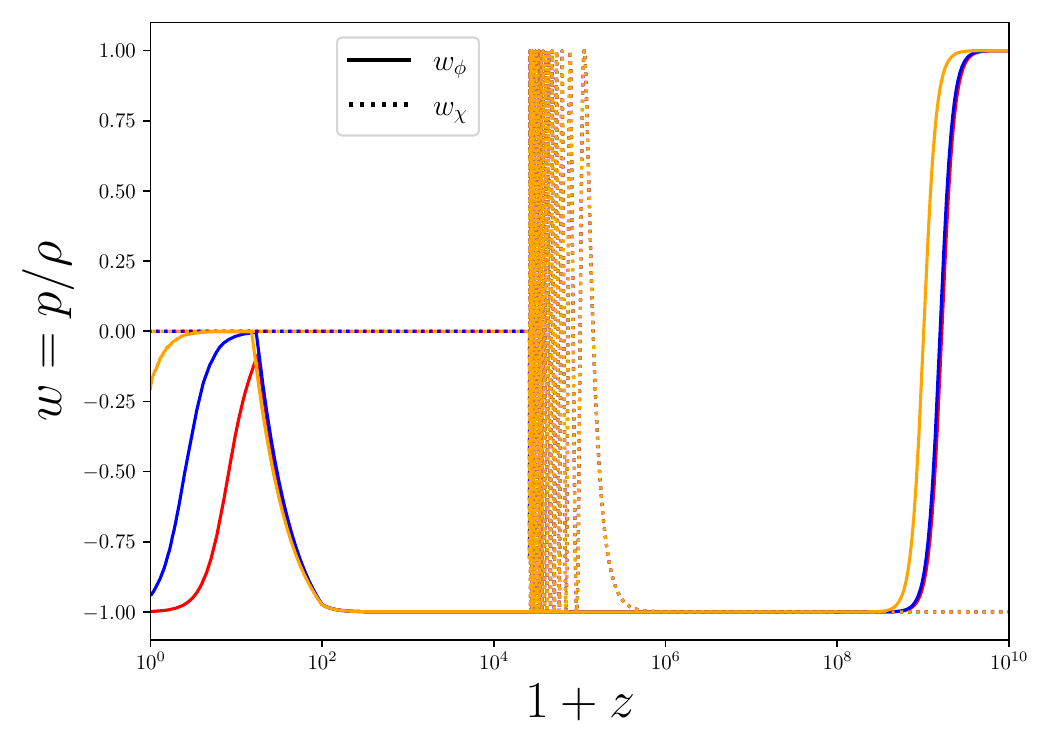}
\caption{Plot of the evolution of the equations of state (EoS) of DM and DE (left) and a time-average over the oscillations in $w_\phi$ (right panel). The DM EoS, $w_\chi$, is being averaged over in both panels after a short period of oscillations. The color code correspond to different values of $\phi_{\rm ini}$ and $\mu^4$ in the presence of DM-DE interaction, expressed in standard \code{CLASS} units of $m_{\rm pl}$ and $m_{\rm pl}^2/\mathrm{Mpc}^2$, respectively. Figure is adapted from ref.~\cite{Aboubrahim:2024cyk}.}
\label{fig1}
\end{centering}
\end{figure}

Depending on the size of $\lambda$, we can distinguish between two regimes: the strong coupling regime and the weak coupling regime. Figure~\ref{fig1} shows the evolution of DM (dotted line) and DE (solid line) equations of state as a function of the redshift $1+z$ for the strong coupling regime, where we have set $\lambda=10\,\,m_{\rm pl}^{-2}\text{Mpc}^{-2}$. Examining the left panel: initially $w_\chi=-1$ and $w_\phi=+1$, so $\chi$ behaves as an early DE component while $\phi$ acts as radiation. Rapid oscillations in $\chi$ begin near $z\sim 10^{5}$, and their averaging yields $w_\chi=0$, causing $\rho_\chi\propto a^{-3}$ as for CDM. The field $\phi$ transitions from radiation-like behavior to $w_\phi\to -1$ as its potential rolls to the minimum and later begins oscillating around $z\sim 10^{2}$. Unlike $\chi$, whose oscillation amplitude is constant, the $\phi$ oscillations decay. For $\phi/F\ll 1$, Eq.~(\ref{kgp0}) becomes
\begin{equation}
\phi^{\prime\prime}+2\mathcal{H}\phi^\prime+a^2m_\phi^2\,\phi\approx 0\,,
\end{equation}
with $m_\phi^2=\lambda\chi^2-\mu^4/F^2$. When $\lambda$ is large enough for $m_\phi^2>0$, oscillations begin once $\mathcal{H}/m_\phi\ll 1$.

Averaging these oscillations, the right panel of Fig.~\ref{fig1} shows $w_\phi$ deviating from and then returning toward $-1$. For $\phi_{\rm ini}=0.1$, $w_\phi$ evolves to $-1$ as expected; for $\phi_{\rm ini}=0.5$, it approaches $w_\phi\sim -0.9$; and for $\phi_{\rm ini}=1.0$, oscillations persist till $z=0$, giving an average EoS far from $-1$, making this value of $\phi_{\rm ini}$ observationally disfavored.

\begin{figure}[H]
\begin{centering}
\includegraphics[width=0.6\textwidth]{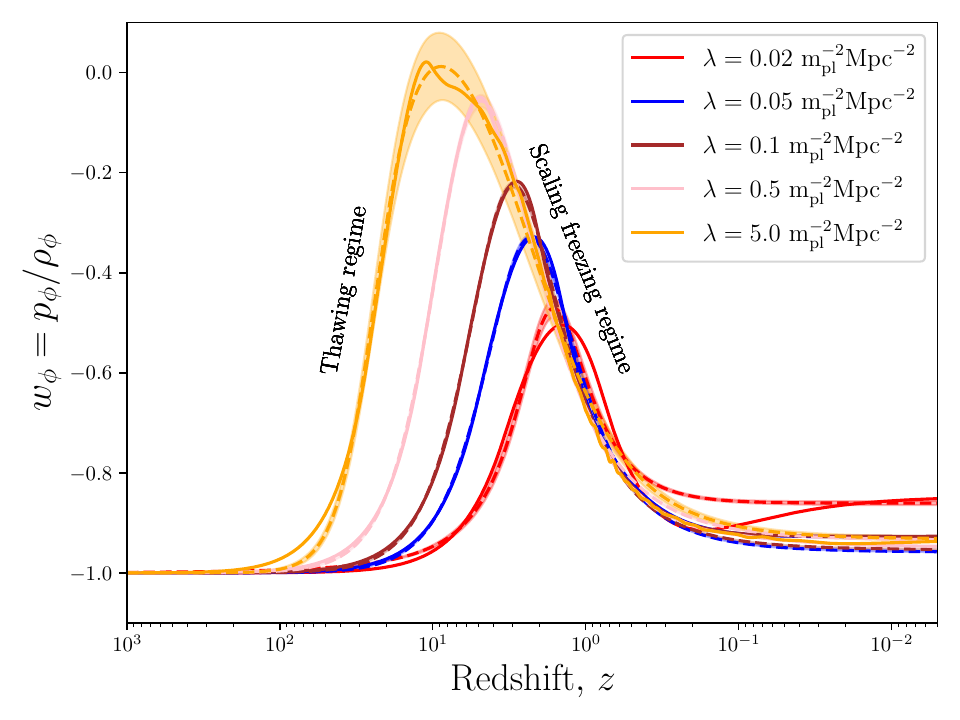}
\caption{The evolution of $w_\phi$ as a function of the redshift $z$ is shown for five choices of the interaction strength $\lambda$. The solid curves display the QCDM predictions, while the dashed curves provide fits using Eq.~(\ref{fit}) together with the corresponding $1\sigma$ uncertainty bands. For all parameter choices, the DE equation of state exhibits a transition induced by $\lambda$, evolving from a thawing behavior at higher redshift to a scaling-freezing behavior at lower redshift. Figure is adapted from ref.~\cite{Aboubrahim:2024cyk}.}
\label{fig2}
\end{centering}
\end{figure}

Figure~\ref{fig2} illustrates the late-time behavior of the DE equation of state, focusing on the peculiar evolution of $w_\phi$ for $z<10^{2}$. The solid curves show the QCDM predictions for five values of the interaction strength $\lambda$, while the dashed curves represent the fit
\begin{equation}
    w(a)=-1+\frac{\alpha\, a^p e^{-p\,a}}{1+(\beta a)^q}\,,
   \label{fit}
\end{equation}
with $\alpha$, $\beta$, $p$, and $q$ as free parameters. For these benchmarks, $w_\phi$ departs from $-1$ in the range $10<z<100$, initially following a thawing-quintessence behavior before turning around and returning toward $-1$ at late times, similar to scaling freezing. Thus, the DM-DE interaction induces a transition of the DE from thawing to scaling freezing, a behavior disfavored by recent DESI constraints.

\begin{figure}[H]
\begin{centering}
\includegraphics[width=0.6\textwidth]{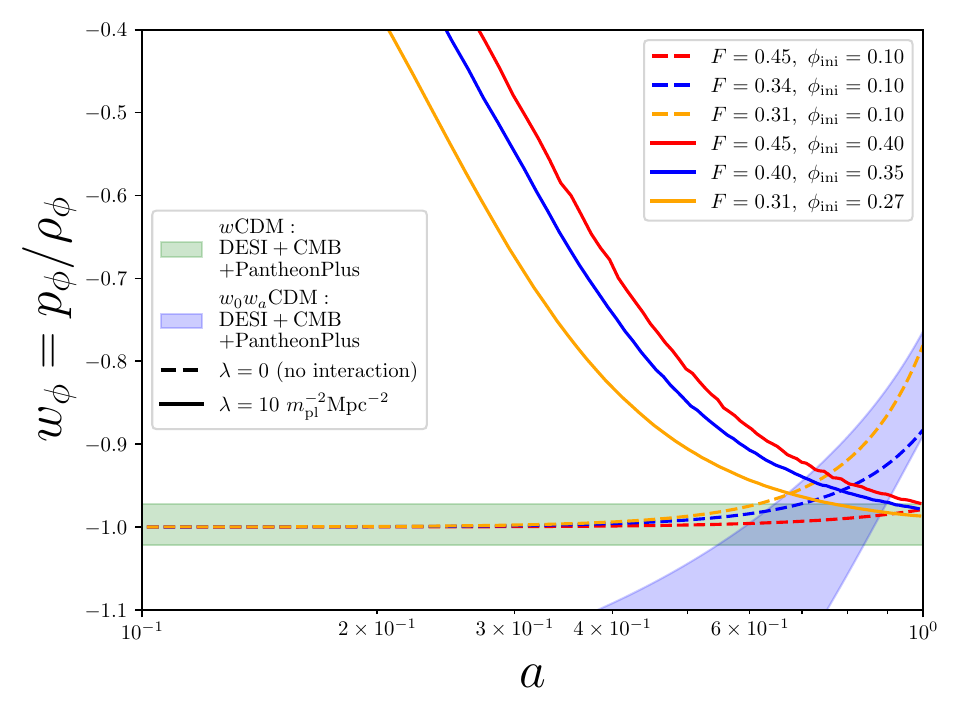}
\caption{The evolution of the DE equation of state in the narrow interval $0.1<a<1$ for the case of no interaction (dashed curves) and the case of DM-DE interaction (solid curves). The colors of each curve correspond to a choice of $(F,\phi_{\rm ini})$ as indicated in the figure legend. Both $F$ and $\phi_{\rm ini}$ are in units of $m_{\rm pl}$. The blue and green bands are the $1\sigma$ regions from DESI's interpretations of their data based on the $w_0 w_a$CDM and $w$CDM models, respectively. Figure is adapted from ref.~\cite{Aboubrahim:2024cyk}.}
\label{fig3}
\end{centering}
\end{figure}

To assess this, Fig.~\ref{fig3} compares two cases: no interaction (dashed curves) and a non-zero interaction (solid curves). The evolution of $w_\phi$ over $0.1<a<1$ is shown along with the DESI+$\mathrm{CMB}$+PantheonPlus $1\sigma$ bounds for the $w$CDM model with constant $w$ (green band) and the CPL form $w(a)=w_0+(1-a)w_a$ known as the $w_0w_a$CDM model (blue band). Without interaction, the EoS stays near $-1$ until $a\sim 0.3$, consistent with thawing PNGB potentials~\cite{Frieman:1995pm}. One benchmark agrees with DESI’s $w$CDM constraints\footnote{See recent analyses in Refs.~\cite{Tada:2024znt,Reboucas:2024smm,Berghaus:2024kra,Wolf:2024stt,Gialamas:2024lyw}.}, while others deviate significantly and align with $w_0w_a$CDM near $a=1$.

When the DM-DE interaction is included, the EoS qualitatively changes: $w_\phi$ evolves toward $-1$ in a manner characteristic of scaling freezing models~\cite{Ferreira:1997au,Copeland:1997et}. The interaction effectively reshapes the DE potential so that the resulting behavior resembles that of a double-exponential form, $V_S(\phi)=\tilde{V}_0(e^{-\lambda_1\phi}+e^{-\lambda_2\phi})$. Both thawing and scaling-freezing regimes for $w_\phi$ are visible in the right panel of Fig.~\ref{fig1}, with a transition occurring near $z_t\sim 10$ when $\lambda\neq 0$.

Using DESI DR1 and DR2 constraints, a quintessence model exhibiting a freezing behavior is incompatible with the allowed ranges of $w_0$ and $w_a$. Although our framework contains several free parameters ($F$, $\phi_{\rm ini}$, $\mu^4$, and $\lambda$), we verified that no parameter choice in the strong coupling regime can reproduce the DESI preferred values. Expanding Eq.~(\ref{fit}) around $\epsilon=(1-a)$ to linear order yields
\begin{align}
  \label{w0}
   w_0&=-1+\frac{\alpha e^{-p}}{1+\beta^q}\,, \\
   w_a&=\frac{q\,\beta^q}{1+\beta^q}(1+w_0)\,,
   \label{wa}
\end{align}
showing that DESI effectively fixes two of the four parameters in Eq.~(\ref{fit}). For instance, one may use measured $w_0$ and $w_a$ to solve Eq.~(\ref{wa}) for $\beta^q$ and Eq.~(\ref{w0}) for $e^{-p}$, leaving $\alpha$ and $q$ free. Using these remaining parameters, we attempted to fit the model predictions but found no viable fit in the strong coupling regime. As anticipated, current DESI data does not support quintessence models with a scaling-freezing equation of state. Therefore, the strong coupling regime is not favored. 

Next, we examine the weak coupling regime, i.e., $\lambda\leq 10^{-2}$. We find that in this regime, the dark energy EoS can be fitted by the function
\begin{align}
    \hat{w}(a)&=-1+\alpha\,e^{-\beta\,a}\arctan(p\,a^q)\,,
    \label{fit-weak}
\end{align}
with $\alpha$, $\beta$, $p$ and $q$ being the fit parameters. Expanding Eq.~(\ref{fit-weak}) up to first order in $(1-a)$, we get $w_0$ and $w_a$ so that
\begin{align}
\label{w0-weak}
    w_0&=-1+\alpha\,e^{\beta}\,\arctan(p)\,, \\
    w_a&=\frac{\alpha\,e^{-\beta}}{1+p^2}\Big[-p\,q+\beta(1+p^2)\arctan(p)\Big]\,.
    \label{wa-weak}
\end{align}
We fit the DE equation of state, $w_\phi(a)$, obtained from our model to the parametrization $\hat{w}(a)=w_0+(1-a)w_a$ and extract the corresponding values of $w_0$ and $w_a$ for each model data point. Among all points, we retain the ones that minimize the $\chi^2$ statistic,
\begin{equation}
    \chi^2=\sum_i\left(\frac{w_i(a)-\hat{w}_i(a)}{\sigma_i}\right)^2,
\end{equation}
which identifies the best-fit model. 

Our model contains four input parameters, and to identify which of them most strongly drives the values of $w_0$ and $w_a$ toward the DESI-preferred $2\sigma$ region, we employ \code{SHAP} (SHapley Additive exPlanations)~\cite{Lundberg:2017uca}. We train a gradient-boosted decision tree classifier (\code{XGBClassifier}) using the \code{XGBoost} framework~\cite{Chen:2016btl} on the Monte Carlo samples generated from our model that represent the best-fit model points. The classifier learns how combinations of input parameters lead to points that lie within the $2\sigma$ constraints on $(w_0, w_a)$. 

\code{XGBoost} produces an ensemble of decision trees that assign probabilities indicating whether a given point satisfies the observational bounds. These probabilities are then analyzed with \code{TreeSHAP}, a fast algorithm within \code{SHAP} that computes exact Shapley values. Treating each model parameter as a feature, the resulting \code{SHAP} values quantify the contribution of each feature to the model outcome, allowing us to determine which parameter most significantly influences whether the predicted $(w_0, w_a)$ fall within DESI limits. Using this techniques, we derive an upper limit on the DM-DE interaction strength, $\lambda$, for each data set. We find that for $\mathrm{DESI+CMB+PantheonPlus}$, $\log\lambda<-2.78$, for $\mathrm{DESI+CMB+Union3}$, $\log\lambda<-2.99$ and for $\mathrm{DESI+CMB+DESY5}$, $\log\lambda<-3.26$. So, the PantheonPlus data set imposes the strongest upper limit on the interaction strength.

\begin{figure}[H]
\begin{centering}
\includegraphics[width=0.49\linewidth]{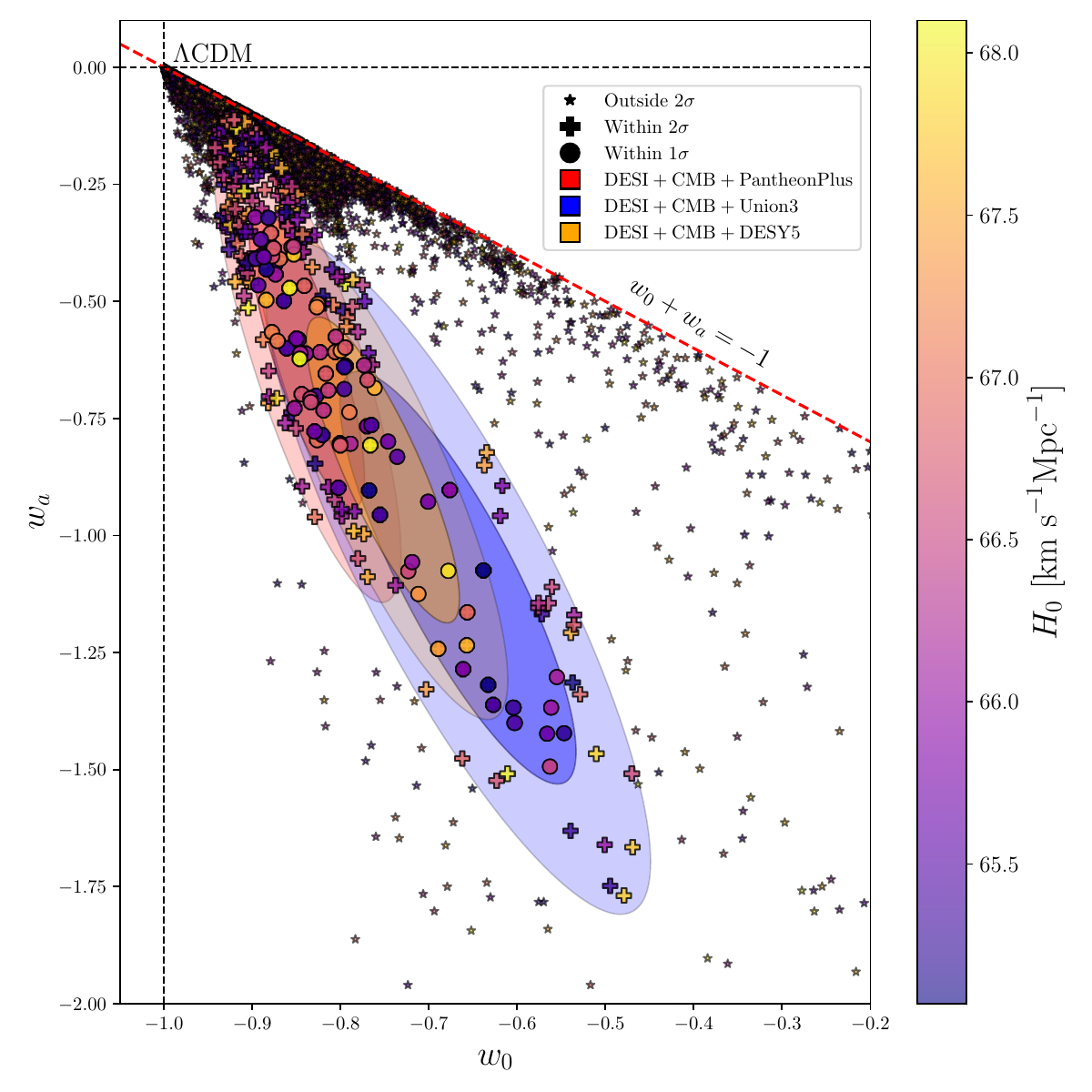}
\includegraphics[width=0.49\textwidth]{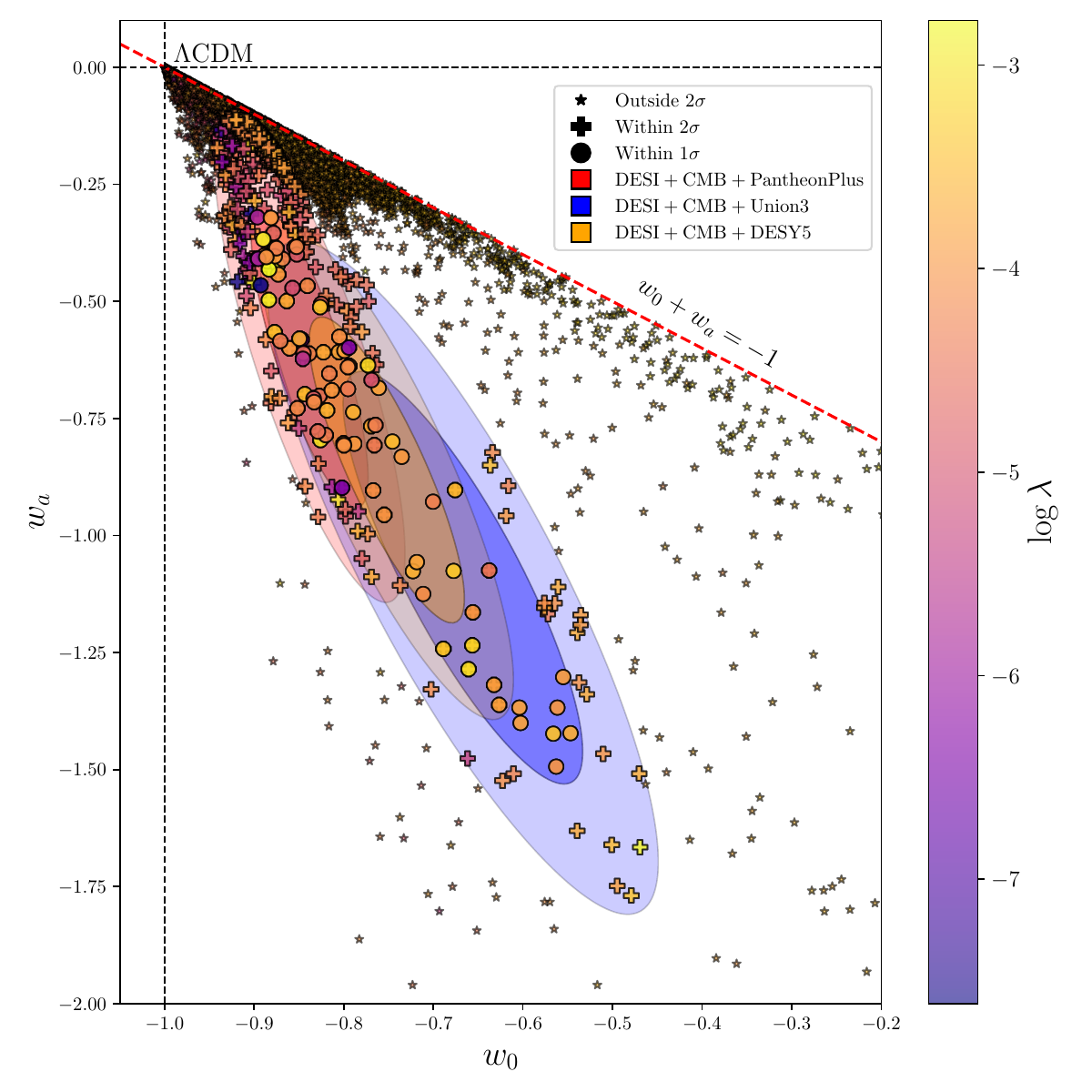}
\caption{Scatter plots in the $w_0$-$w_a$ plane are shown overlaid on the DESI DR2 posterior contours with the upper bound on $\log\lambda$ imposed for each data set. The left panel uses the Hubble parameter $H_0$ as the color scale, while the right panel displays $\log\lambda$. The model yields parameter points that fall within the $1\sigma$ and $2\sigma$ posterior regions of the considered data sets. Figure is adapted from ref.~\cite{Aboubrahim:2024cyk}.}
\label{fig4}
\end{centering}
\end{figure}

Figure~\ref{fig4} displays the model points that satisfy the imposed upper bound on $\lambda$ as well as the constraints on $H_0$ and $\Omega_{\rm m}$. Even with $\lambda$ values approaching the upper limit, a substantial number of points continue to fall within the $1\sigma$ and $2\sigma$ posterior regions of DESI's data set combination in the $w_0$-$w_a$ plane.

\section{Extraction of cosmological parameters with MCMC and Bayesian inference}

Because a DM-DE interaction affects not only the background evolution but also density perturbations (and hence the CMB temperature/polarization spectra and the matter power spectrum), we confront the model with current cosmological data: Planck 2018 CMB measurements, DESI-DR2 BAO, and several SN-based local distance data sets. We implement the model in \code{CLASS} by evolving the background equations together with the synchronous-gauge KG perturbation equations for the fields $\phi$ and $\chi$, namely Eqs.~(\ref{kgp1}) and~(\ref{kgc1}), while all other species follow their standard $\Lambda$CDM evolution. We set $\chi_1=\chi_1^\prime=0$ and $\phi_1=\phi_1^\prime=0$ initially (analogous to $\delta_{\rm ini}=0$ and $\Theta_{\rm ini}=0$), noting that the perturbations are rapidly driven to the attractor solution~\cite{Ballesteros:2010ks}. Since the synchronous gauge is not fully fixed by a scalar-field DM component (because $w_\chi$ is dynamical rather than identically zero), we retain the use of synchronous gauge in \code{CLASS} by including a small CDM component, $\Omega_{\rm CDM}h^2=10^{-5}$.

We use the following likelihoods and data sets: \textbf{CMB} (Planck TT, TE, EE, low-$\ell$, and lensing)~\cite{Planck:2018vyg,Planck:2018nkj,Planck:2019nip,Planck:2018lbu}, \textbf{DESI DR2} (BAO measurements)~\cite{DESI:2025zgx,DESI:2025zpo}, \textbf{PPS} (PantheonPlus+SH0ES)~\cite{Brout:2022vxf,Riess:2021jrx}, \textbf{Union3} (2087 SN Ia)~\cite{Rubin:2023jdq}, and \textbf{DESY5} (DES Year 5 SN sample)~\cite{DES:2024jxu}. 

For inference, we interface \code{CLASS} with \code{Cobaya}~\cite{Torrado:2020dgo} to perform MCMC sampling (adaptive, speed-hierarchy-aware sampler adapted from \code{CosmoMC})~\cite{Lewis:2002ah,Lewis:2013hha} using fast-dragging~\cite{Neal:2005uqf}. Convergence is assessed with the Gelman-Rubin criterion $R-1<0.05$~\cite{Gelman:1992zz}, and posterior constraints are extracted with \code{GetDist}~\cite{Lewis:2019xzd}. The sampled parameters include the six $\Lambda$CDM parameters plus $\log\mu^4$, $F$, $\phi_{\rm ini}$, and $\log\lambda$, all with flat priors.

Including perturbations, the results (Table~\ref{tab1}) provide 68\% CL constraints for multiple data combinations. The baryon density fraction remains nearly unchanged across combinations, while the inferred DM fraction increases by more than $\sim 5\%$ when SN data such as DESY5 and Union3 are included relative to the first two combinations. The model parameter $\mu^4$ is constrained to values of order $10^{-8}~\mathrm{m}_{\rm Pl}^2/\mathrm{Mpc}^2$ (e.g. from $\sim 5.5\times 10^{-8}$ for CMB+DESI up to $\sim 7.6\times 10^{-8}$ for CMB+DESI+Union3). The tightest lower bound on $F$ arises from CMB+DESI+PPS (e.g. $F>0.62$), while it relaxes with DESY5 inclusion (down to $F\sim 0.29$). An upper bound on $\phi_{\rm ini}$ is obtained, with PPS giving the most stringent limit. All data combinations favor weak coupling, yielding upper limits in the range $\lambda\lesssim 10^{-3}$ down to $\lambda\lesssim 10^{-5.7}$, which are stronger than constraints obtained from background evolution alone. The model only modestly alleviates the Hubble tension at 95\% CL (largest shifts driven by PPS). For structure growth, the derived $S_8\equiv \sigma_8\sqrt{\Omega_{\rm m}/0.3}$ is consistent with recent \code{KiDS-Legacy} results~\cite{Wright:2025xka}. Finally, using Eqs.~(\ref{w0-weak}) and~(\ref{wa-weak}) with DESI DR2, the inferred values remain close to $\Lambda$CDM, $w_0\simeq -1$ and $w_a\simeq 0$, although the SN data (notably DESY5 and Union3) significantly shifts the preferred $(w_0,w_a)$ and enlarges evidence for mild evolution.

\begin{table}[H]
\centering
{\tabulinesep=1.2mm
\resizebox{\textwidth}{!}{\begin{tabu}{ccccc}
\hline\hline
\textbf{Parameter} & \textbf{CMB+DESI} & \textbf{CMB+DESI+PPS} & \textbf{CMB+DESI+DESY5} & \textbf{CMB+DESI+Union3} \\
\hline
{$\log(10^{10} A_\mathrm{s})$} & $3.042\pm 0.010            $ & $3.046\pm 0.013            $ & $3.044\pm 0.011            $ & $3.045\pm 0.012            $\\
{$n_\mathrm{s}   $} & $0.9684\pm 0.0036          $ & $0.9703\pm 0.0033          $ & $0.9696\pm 0.0037          $ & $0.9694\pm 0.0035          $\\
{$100\theta_\mathrm{s}$} & $1.04217\pm 0.00029        $ & $1.04227\pm 0.00027        $ & $1.04219\pm 0.00028        $ & $1.04220\pm 0.00028        $\\
{$\Omega_\mathrm{b} h^2$} & $0.02278\pm 0.00011        $ & $0.02283\pm 0.00011        $ & $0.02278\pm 0.00011        $ & $0.02279\pm 0.00011        $\\
{$\Omega_\chi    $} & $0.2420\pm 0.0083          $ & $0.2415\pm 0.0066          $ & $0.256^{+0.011}_{-0.0097}  $ & $0.2543^{+0.0093}_{-0.012} $\\
{$z_\mathrm{reio}$} & $7.69\pm 0.48              $ & $7.86\pm 0.61              $ & $7.80\pm 0.56              $ & $7.84\pm 0.60              $\\
{$\log\mu^4      $} & $-7.233^{+0.013}_{-0.028}  $ & $-7.230^{+0.013}_{-0.024}  $ & $-7.192^{+0.041}_{-0.065}  $ & $-7.175^{+0.053}_{-0.075}  $\\
{$F~[m_{\rm Pl}]$} & $> 0.561                   $ & $> 0.620                   $ & $0.51^{+0.20}_{-0.22}      $ & $0.63^{+0.23}_{-0.17}      $\\
{$\phi_{\mathrm{ini}}~[m_{\rm Pl}]$} & $< 0.157                   $ & $< 0.263                   $ & $< 0.384                   $ & $< 0.560                   $\\
{$\log\lambda    $} & $-4.9^{+2.0}_{-2.7}        $ & $-5.49^{+0.97}_{-2.2}      $ & $< -5.49                   $ & $< -5.69                   $\\
\hline\hline
$H_0~[\rm km/s/Mpc]$ & $69.4^{+1.1}_{-0.98}       $ & $69.39\pm 0.87             $ & $67.6^{+1.3}_{-1.6}        $ & $67.8^{+1.6}_{-1.3}        $\\
$\Omega_\mathrm{m}         $ & $0.276^{+0.029}_{-0.016}   $ & $0.280^{+0.025}_{-0.011}   $ & $0.294^{+0.028}_{-0.015}   $ & $0.295^{+0.024}_{-0.015}   $\\
$\Omega_\phi               $ & $0.7092\pm 0.0097          $ & $0.7097^{+0.0079}_{-0.0071}$ & $0.693\pm 0.012            $ & $0.695^{+0.014}_{-0.011}   $\\
$\tau_\mathrm{reio}          $ & $0.0559\pm 0.0049          $ & $0.0579^{+0.0060}_{-0.0068}$ & $0.0571^{+0.0054}_{-0.0061}$ & $0.0575\pm 0.0062          $\\
$S_8                       $ & $0.769^{+0.034}_{-0.018}   $ & $0.773^{+0.028}_{-0.014}   $ & $0.778^{+0.027}_{-0.014}   $ & $0.780^{+0.023}_{-0.012}   $\\
$w_0                       $ & $-0.900^{+0.069}_{-0.25}   $ & $-0.937^{+0.021}_{-0.17}   $ & $-0.925^{+0.055}_{-0.13}   $ & $-0.890^{+0.075}_{-0.16}   $\\
$w_a                       $ & $-0.018^{+0.035}_{-0.016}  $ & $-0.018^{+0.029}_{-0.013}  $ & $-0.076^{+0.11}_{-0.046}   $ & $-0.109^{+0.14}_{-0.070}   $\\
\hline
$\Delta\chi^2_{\rm min}$ & 25.11 & 21.61 & 14.23 & 21.05 \\
$\Delta\mathrm{AIC}$ & 33.11 & 29.61 & 22.23 & 29.05 \\
\hline\hline
\end{tabu}}}
\caption{Constraints on some of the cosmological parameters of our model. The values are quoted at 68\% CL intervals for three data set combinations. The middle double line separates the sampled and derived parameters using MCMC. In the last two rows we show the values of $\Delta\text{AIC}\equiv\text{AIC}_{\rm QCDM}-\text{AIC}_{\Lambda\rm CDM}$ and $\Delta\chi^2_{\rm min}\equiv\chi^2_{\rm QCDM,min}-\chi^2_{\rm \Lambda CDM,min}$.  Table is taken from ref.~\cite{Aboubrahim:2024cyk}. }
\label{tab1}
\end{table}

To compare QCDM against $\Lambda$CDM, we compute the Akaike Information Criterion (AIC)~\cite{Akaike:1974vps},
\begin{equation}
\mathrm{AIC}\equiv -2\ln\mathcal{L}_{\rm max}+2K,
\end{equation}
and define $\Delta\mathrm{AIC}\equiv \mathrm{AIC}_{\rm QCDM}-\mathrm{AIC}_{\Lambda\mathrm{CDM}}$. Using the usual rule of thumb ($\Delta\mathrm{AIC}<-5$ favors QCDM, while $\Delta\mathrm{AIC}>10$ decisively favors $\Lambda$CDM), the results in Table~\ref{tab1} indicate $\Delta\mathrm{AIC}>10$ for all data combinations, i.e. the data prefer $\Lambda$CDM over the interacting QCDM model.

\begin{figure}[H]
\begin{centering}
\includegraphics[width=0.49\linewidth]{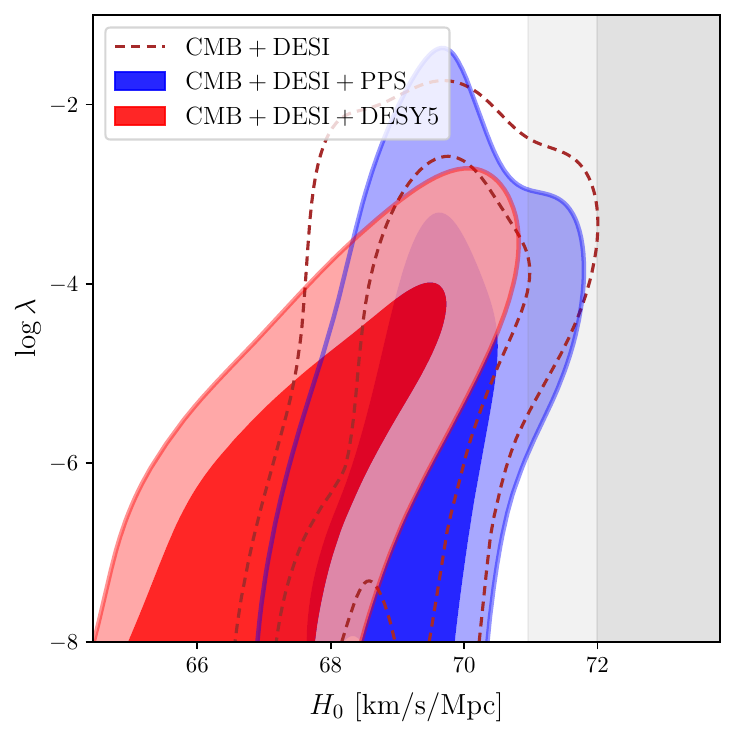}
\includegraphics[width=0.49\textwidth]{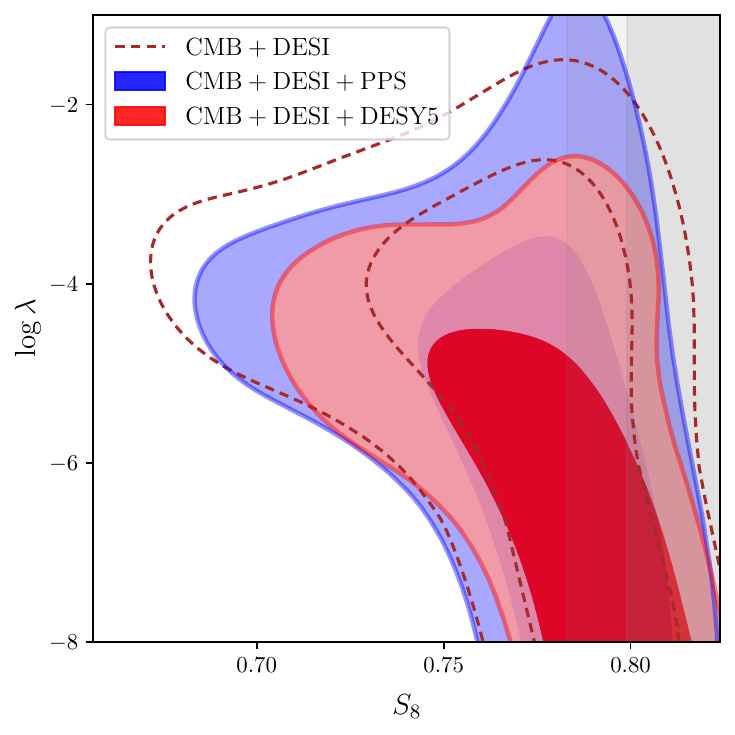}
\caption{The two-dimensional marginalized contours at the 68\% and 95\% confidence levels are shown for the DM-DE coupling strength versus $H_0$ (left panel) and versus $S_8$ (right panel). In the left panel, the gray bands indicate the SH0ES~\cite{Riess:2021jrx} measurement, $H_0=73.04\pm 1.04~\mathrm{km\,s^{-1}\,Mpc^{-1}}$, while in the right panel they represent the \code{KiDS}-Legacy~\cite{Wright:2025xka} constraint, $S_8=0.815^{+0.016}_{-0.021}$. Figure is adapted from ref.~\cite{Aboubrahim:2024cyk}. }
\label{fig5}
\end{centering}
\end{figure}

Figure~\ref{fig5} shows the correlation between $H_0$ and $\log\lambda$ (left) and between $S_8$ and $\log\lambda$ (right), with the gray bands indicating the $1\sigma$ and $2\sigma$ experimental ranges. The $1\sigma$ contours of $H_0$ for both data sets do not overlap with the measured value of $H_0$, whereas the corresponding contours for $S_8$ lie within its experimental band. Thus, the QCDM model only mildly alleviates the $H_0$ tension, while $S_8$ shows no significant tension.

\begin{figure}[H]
\begin{centering}
\includegraphics[width=0.49\linewidth]{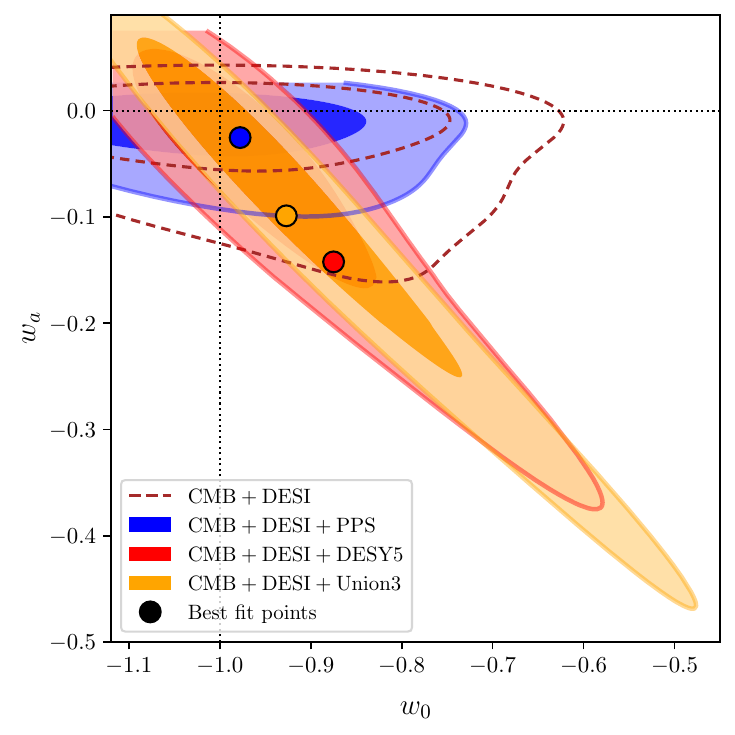}
\includegraphics[width=0.49\textwidth]{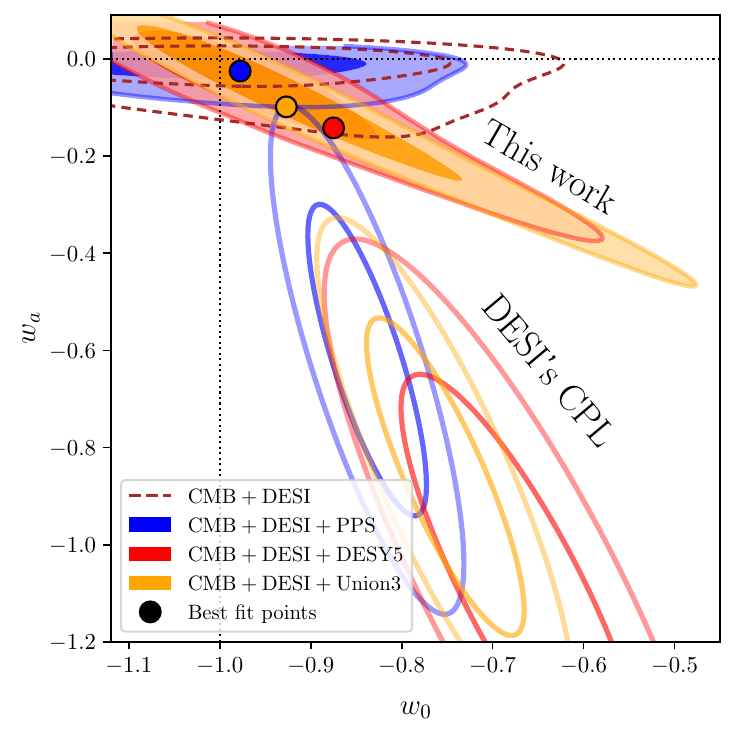}
\caption{Left panel: Posterior contours for $(w_0,w_a)$ obtained from the four data-set combinations: CMB+DESI (dashed), CMB+DESI+PPS (blue), CMB+DESI+DESY5 (red), and CMB+DESI+Union3 (orange). The dotted vertical and horizontal lines mark the $\Lambda$CDM point $(w_0,w_a)=(-1,0)$. Filled dots indicate the best-fit values for the PPS, DESY5, and Union3 cases. Right panel: Same as the left panel, with DESI's unfilled contours added for comparison. Figure is adapted from ref.~\cite{Aboubrahim:2024cyk}. }
\label{fig6}
\end{centering}
\end{figure}

We now examine how the different data sets impact the inferred values of $w_0$ and $w_a$, and what this implies for evolving dark energy. The left panel of Fig.~\ref{fig6} shows the $1\sigma$ and $2\sigma$ contours in the $w_0$-$w_a$ plane for the four data set combinations. A substantial portion of the allowed region lies in the fourth quadrant ($w_0<0$, $w_a<0$); however, unlike the DESI DR2 results, parts of the contours extend into other quadrants and overlap with the $\Lambda$CDM point $(w_0,w_a)=(-1,0)$. Although the constraints remain consistent with $w_0=-1$, the $1\sigma$ and $2\sigma$ regions still leave ample room for an evolving DE equation of state near $a=1$. The best-fit points, marked by filled dots for each of the PPS, DESY5, and Union3 data sets, all fall in the fourth quadrant, thereby favoring evolution in the DE component. 

In contrast to PantheonPlus+SH0ES, the DESY5 and Union3 samples push the preferred values of $(w_0,w_a)$ toward an evolving DE scenario while staying within the quintessence regime (i.e., without crossing the phantom divide). It is important to emphasize that although we describe the DE EoS using the CPL form, $w_0$ and $w_a$ are \emph{derived} quantities in this model and depend on the underlying particle physics parameters rather than being freely varied. To highlight the difference from DESI’s analysis, we overlay DESI’s (unfilled) contours in the right panel of Fig.~\ref{fig6}. These contours reach further into the fourth quadrant relative to our results, reflecting the phantom-divide crossing behavior that appears when CPL is treated as a purely phenomenological parametrization.

\section{Conclusion}

In this work, we investigated a field-theoretic cosmological model in which dark matter and dark energy are described by ultralight spin zero fields coupled through an explicit interaction term.  Using a specific choice of dark energy potential together with the interaction term, we analyzed the resulting evolution of the dark energy equation of state.

Our study identifies two qualitatively distinct regimes: a strong-coupling region ($\lambda \geq 10^{-2}$) and a weak-coupling region ($\lambda < 10^{-2}$). In the strong-coupling regime, the DE equation of state undergoes a transmutation from thawing behavior at early times to scaling-freezing behavior at late times. Because a freezing DE equation of state today is disfavored by DESI, this regime is effectively ruled out. In contrast, the weak-coupling regime yields evolution consistent with DESI, supporting a mildly evolving DE component. We provided a fit for $w_\phi(a)$ which, to leading order in $(1-a)$, produces values of $(w_0,w_a)$ compatible with DESI without requiring a phantom-divide crossing.

Using DESI data set combinations, we obtained an upper limit on $\lambda$ and showed that many benchmark points near this limit remain viable. Incorporating cosmological perturbations allowed us to derive tighter constraints on the full parameter space. Overall, the results indicate that the QCDM framework accommodates a cosmological-constant-like behavior while still permitting an evolving DE equation of state. This is reflected in the best-fit points across all data sets, which fall in the fourth quadrant of the $(w_0,w_a)$ plane, signaling a preference for evolving quintessence.

\acknowledgments
The research of PN was supported in part by the NSF Grant PHY-2209903. The analysis presented here was done using the computing resources of the Phage Cluster at Union College.

\bibliographystyle{JHEP}
\bibliography{references}

\end{document}